\documentclass[conference]{IEEEtran}

\usepackage{amsthm,amssymb,graphicx,multirow,amsmath,color,amsfonts,subfig,balance}
\usepackage{textcomp}
\usepackage[latin1]{inputenc}
\usepackage[noadjust]{cite}
\usepackage{tikz}
\usepackage{fontenc}

\usepackage[latin1]{inputenc}
\usepackage{tikz}
\usetikzlibrary{shapes,arrows}
\usepackage{bbm} 
\usepackage{graphicx} 
\usepackage{pdfpages}

\usepackage{tabulary}
\usepackage{multirow}
\usepackage{comment}
\usepackage{algpseudocode}
\usepackage{todonotes}
\def\beq{\begin{equation}}
\def\eeq{\end{equation}}
\def\beqa{\begin{eqnarray}}
\def\eeqa{\end{eqnarray}}
\def\beqan{\begin{eqnarray*}}
\def\eeqan{\end{eqnarray*}}

\include{notation_new}
\DeclareMathOperator*{\argmin}{argmin} 
\makeatletter
\def\bstctlcite{\@ifnextchar[{\@bstctlcite}{\@bstctlcite[@auxout]}}
\def\@bstctlcite[#1]#2{\@bsphack
  \@for\@citeb:=#2\do{%
    \edef\@citeb{\expandafter\@firstofone\@citeb}%
    \if@filesw\immediate\write\csname #1\endcsname{\string\citation{\@citeb}}\fi}%
  \@esphack}
\makeatother

\begin{document}

\title{Beamformed mmWave System Propagation at 60 GHz in an Office Environment}
\author{Syed Hashim Ali Shah$^{*}$, Sarankumar Balakrishnan \textsuperscript{\textdagger}, Liangxiao Xin\textsuperscript{\textdaggerdbl}, Mohamed Abouelseoud\textsuperscript{\textdaggerdbl}, \\
Kazuyuki Sakoda\textsuperscript{**} , Ken Tanaka\textsuperscript{**} , Christopher Slezak$^{*}$, Sundeep Rangan$^{*}$ and Shivendra Panwar$^{*}$ \\
$^{*}$NYU Wireless, New York University, Brooklyn, NY, USA, \\  \textsuperscript{\textdagger} Dept. of Electrical Engineering, University at Buffalo, NY, USA\\
\textsuperscript{\textdaggerdbl} US Research Center, Sony Corporation of America, CA, USA\\
\textsuperscript{**}Connectivity Technology Development Dept.,Fundamental Technology R\&D Field,\\ R\&D Center, Sony Corporation, Japan
\\
{\footnotesize e-mail: \{s.hashim, chris.slezak, srangan, panwar\}@nyu.edu}
\footnotesize, sarankum@buffalo.edu, \\
\footnotesize \{Kazuyuki.Sakoda, Ken.A.Tanaka, mohamed.abouelseoud\}@sony.com, Liangxiao.Xin@am.sony.com
}

\maketitle
\bstctlcite{IEEEexample:BSTcontrol}
\begin{abstract}

Millimeter wave wireless systems rely
heavily on directional communication in narrow
steerable beams.
Tools to measure the spatial and temporal nature
of the channel are necessary to evaluate 
beamforming and related algorithms. 
This paper presents a novel 60$~$GHz phased-array
based directional channel sounder and
data analysis procedure that can accurately 
extract paths and their transmit and receive directions.
The gains along each path can also be measured for
analyzing blocking scenarios.  
The sounder is validated in an indoor office environment.
\end{abstract}


\section{Introduction}
\label{sec:intro}

Millimeter wave (mmWave) technology is pivotal in designing the future of wireless communication systems due to the massive 
available bandwidth\cite{sundeepnlos}.
To compensate for the high isotropic path loss and enable
spatial multiplexing gains, 
mmWave communication is generally performed in narrow
beams formed by phased arrays \cite{mmwaveitwillwork,beamformingcitation}.
Systems typically exploit both 
line of sight (LOS) and non-line of sight (NLOS) components for propagation between a transmitter (TX) and a receiver (RX) \cite{nlosimportance} \cite{sundeepnlos}. The mmWave beams in any particular direction 
can be blocked easily due to smaller Fresnel zones 
\cite{rappaport2014millimeter} and the penetration loss suffered by mmWave systems
is worse compared to sub 6-GHz systems \cite{PenLoss}.
A key challenge in designing mmWave systems is finding beamforming algorithms
that can identify and switch to the optimal beam 
in environments with multiple paths and blockage.
Proper evaluation of such algorithms relies on tools to 
accurately measure available paths in realistic settings and observe how the gains
and directions of the paths vary due to blockage and motion.

This paper presents a 60 GHz directional channel sounder
using a phased array measurement system in \cite{Slezak201760GB,chrismeasurement}
along with a novel data analysis procedure that can identify the real-time 
directions and gains of path in complex multiple path environments.
The  60 GHz bands are key unlicensed bands  \cite{heath60G} used for both
 802.11ad and 802.11ay \cite{adcitation,aycitation}.
Prior propagation work at 60 GHz has been mostly 
done using quasi-omni antennas.  For example,
\cite{hornantennaref,hornantenna60Ghz} use horn antennas and 
in \cite{JacobosQuasiOmni}, 
ray-tracing of the room was performed since no spatial information was available.
This work uses phased arrays, similar to those used in
 \cite{x60Ref}, to measure the channel in multiple directions simultaneously.
Phased array measurements \cite{keurnermeasurement} were used to estimate
coverage with human blockage. The use of phased arrays provides spatial information about the channel which is of paramount importance in mmWave systems as compared to quasi-omni antennas, which provide very limited spatial information.

In this paper, we use measurements from an office cubicle venue to analyze the behavior of LOS and NLOS components of a mmWave system operating at 60 GHz. The contribution of this paper is a novel
method identifying both LOS and NLOS paths and 
estimating their angles of arrival (AoA)/ angles of departure (AoD) from the performed measurements.  The estimates
are validated against a commercial ray tracer given a detailed interior
model.  
We also perform a preliminary blockage analysis for all the components and observe the evolution of link quality on all components with time. These analyses will aid in design of robust mmWave wireless systems.

\label{sec:motivation}
\begin{figure}[b!]
 \centering
        
       \includegraphics[trim={0cm 0cm 0cm 5cm},clip,width=0.45\textwidth]{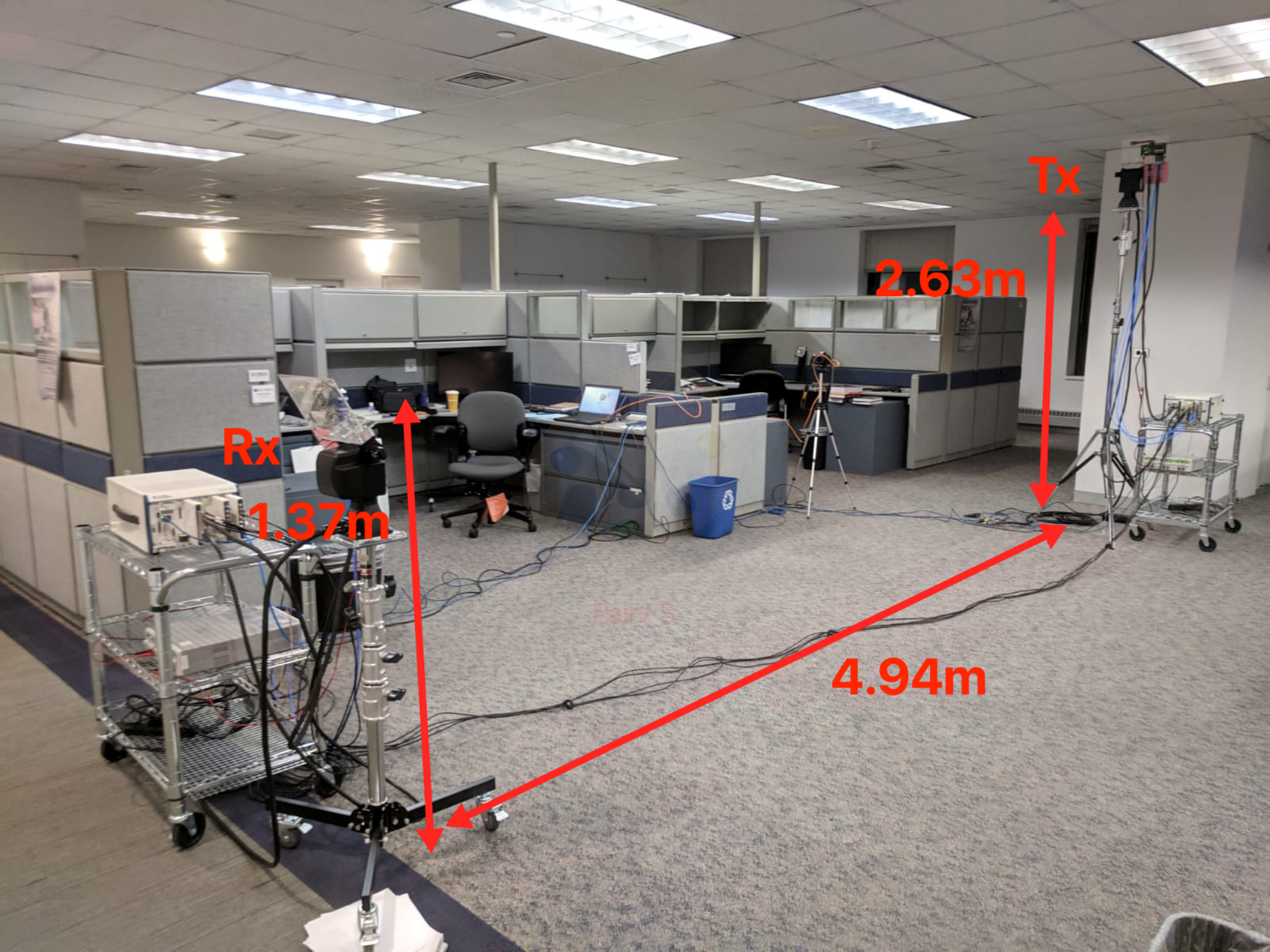}
\caption{Venue for the measurements at NYU Wireless Office space}
\label{fig:venueview}
\end{figure}

\section{Measurement Setup}
\label{sec:MeasurementSetup}

\paragraph{Measurement venue}
The measurement venue is an office space shown in Fig \ref{fig:venueview}, similar venue was used for measurements in \cite{venuecitation}. The TX and RX had a clear LOS path and were both located in a relatively open area within the office. The TX was located at a height of 263 cm (ceiling height of 273 cm) on top of a gimbal and the RX was at a height of 137 cm on a gimbal which in turn is mounted on tripod. This was done to imitate a scenario replicating an access point on the ceiling serving a user hand-held device. 

\begin{figure}[t]

  \begin{minipage}{.25\textwidth}
    \includegraphics[width=.9\textwidth]{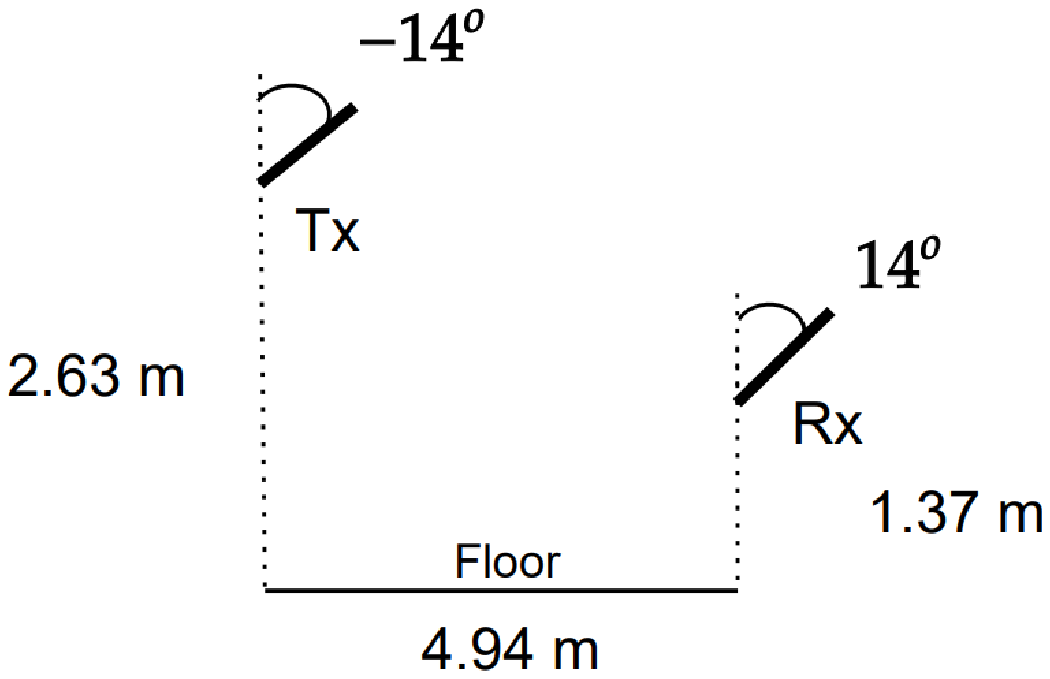}
    \caption{Side-view}
  \end{minipage}%
  \begin{minipage}{.25\textwidth}
    \includegraphics[width=.6\textwidth]{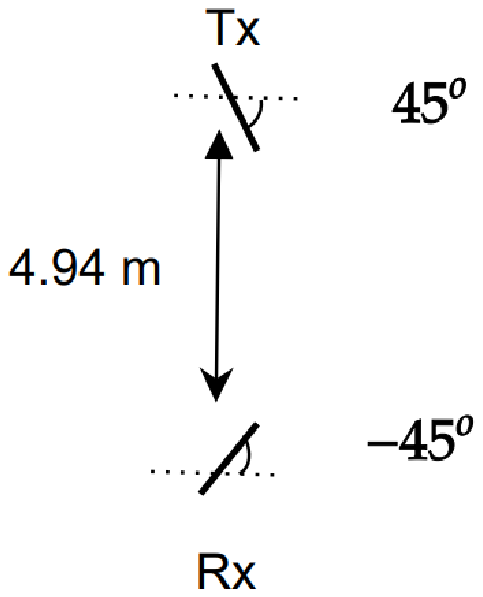}
     \caption{Top-view}
  \end{minipage}
  \caption{Side and top view for TX-RX orientations for case 6 with bold lines showing the array mounts}
   \label{fig:case6orientation}
\end{figure}

Since the arrays on both TX and
RX have limited steerable range,
a total of 12 measurements were 
done, each with different TX and RX orientations, which we refer to as \textit{cases}. The coordinate system for the measurements is described as follows. The rotation(Azimuth) is said to be positive when the gimbal is rotated clockwise as viewed from above with respect to its axis of rotation. It is counter clockwise vice versa. For elevation, an up-tilt from the horizon is considered positive while a down-tilt from the horizon is considered negative.
The orientations for all the cases have been summarized in Table \ref{table:antorientation} according to the described co-ordinate system. We call $(\theta_{TX}^0,\theta_{RX}^0,\phi_{TX}^0,\phi_{RX}^0)$ the reported angles. The top-view and side view for case 6 have been shown in Fig. \ref{fig:case6orientation} to explain the coordinate system.
\begin{table}[!t]
\centering
\begin{tabular}{|l|l|l|l|l|}

\hline
\textbf{Cases} & \textbf{\begin{tabular}[c]{@{}l@{}}TX \\ Elevation\\ $\theta_{TX}^0$\end{tabular}} & \textbf{\begin{tabular}[c]{@{}l@{}}RX\\ Elevation\\ $\theta_{RX}^0$\end{tabular}} & \textbf{\begin{tabular}[c]{@{}l@{}}TX \\ Azimuth\\ $\phi_{TX}^0$\end{tabular}} & \textbf{\begin{tabular}[c]{@{}l@{}}RX\\ Azimuth\\ $\phi_{RX}^0$\end{tabular}} \\ \hline

1              & -14                                                                            & 0                                                                             & -45                                                                        & 45                                                                        \\ \hline
2              & -14                                                                            & 0                                                                             & 0                                                                          & 0                                                                         \\ \hline
3              & -14                                                                            & 0                                                                             & 45                                                                         & -45                                                                       \\ \hline
4              & -14                                                                            & 14                                                                            & -45                                                                        & 45                                                                        \\ \hline
5              & -14                                                                            & 14                                                                            & 0                                                                          & 0                                                                         \\ \hline
6              & -14                                                                            & 14                                                                            & 45                                                                         & -45                                                                       \\ \hline
7              & 0                                                                              & 0                                                                             & -45                                                                        & 45                                                                        \\ \hline
8              & 0                                                                              & 0                                                                             & 0                                                                          & 0                                                                         \\ \hline
9              & 0                                                                              & 0                                                                             & 45                                                                         & -45                                                                       \\ \hline
10             & 0                                                                              & 14                                                                            & -45                                                                        & 45                                                                        \\ \hline
11             & 0                                                                              & 14                                                                            & 0                                                                          & 0                                                                         \\ \hline
12             & 0                                                                              & 14                                                                            & 45                                                                         & -45                                                                       \\ \hline
\end{tabular}
\caption{Array Orientation for all cases}
\label{table:antorientation}
\end{table}

\begin{figure*}[t!]

  \includegraphics[trim={3cm 1.5cm 3cm 0cm},clip,width=\textwidth]{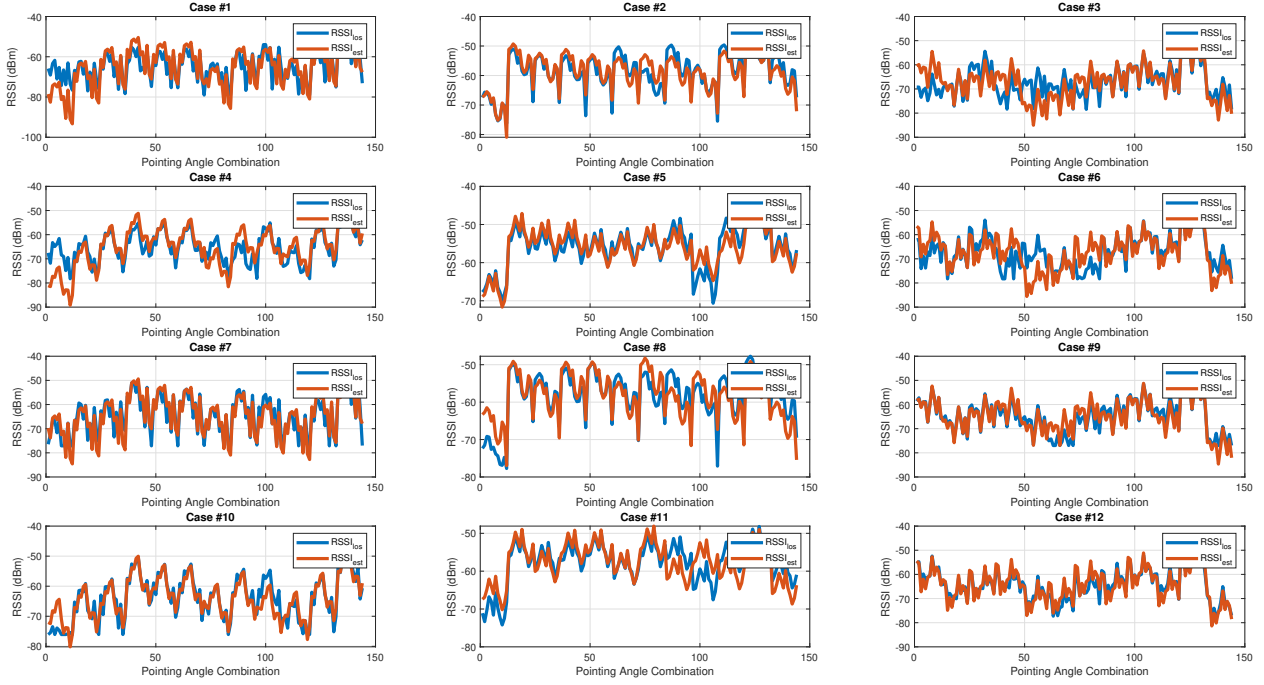}
  \caption{Comparison of $RSSI_{los}$ (blue) vs $RSSI_{est}$(red) for all 12 cases }
 \label{fig:losvalidation}
\end{figure*}

\paragraph{Phased Array Measurements} 
In each case, the channel is measured
using the phased array sounder
described in detail in  
\cite{Slezak201760GB,chrismeasurement}. The sounder has 12 element SiBeam phased arrays at both the TX and RX,
with fixed codebooks of 12 TX and
12 RX directions.  Hence, there
are a total of $N_{dir}=144$ pointing angle combinations (PACs). 
The channel is measured with multiple
\emph{scans}.  In each scan, the
TX and RX sweep all $N_{dir}$ PACs.
Each scan takes $3.2\mathrm{ms}$. There are a total of $N_{scan}=1750$ scans, meaning the total measurement time is 5.6s. Each scan generates 144 power delay profiles (PDPs) containing $N_{dly}=192$ samples for every PAC. All of this data is written into a tensor with dimensions $N_{dly}\times\ N_{dir}\times N_{scan}$.   Sampling rate used for measurements was 1.25 giga samples per second ($GS/s$), so the difference between two samples is 0.8 $ns$. 

\paragraph{Antenna Patterns} For the  SiBeam 60 GHz phased array were already available for TX and RX arrays for all codebooks \cite{x60Ref}. The antenna patterns are organized into four dimensional matrices for TX and RX where each entry corresponds to the directionality $G_{TX/RX}$ in dB, at the given azimuth$(\phi) $ and elevation angle $(\theta) $ for the given direction $c$. We define the total directionality in the $n$-th PAC for antenna as $G(n,\omega)$ where $\omega = (\phi_{TX},\phi_{RX},\theta_{TX},\theta_{RX})$. The $n$-th entry of $G(n,\omega)$ is just the sum of the TX and RX gains (in dB) in the $n$-th PAC for the given AoA/AoD pair $\omega$.

\paragraph{Omni-directional PDP Synthesis} From section \ref{sec:MeasurementSetup}, we acquire 144 PDPs for different beam pairs from each scan in every measurement. These PDPs strongly depend on the beamforming deployed at TX and RX and hence are directional in nature. Each PDP corresponds to a specific beam pair so getting a general mapping of the spatial information is difficult. To get a general idea of the spatio-temporal information of the channel, these directional PDPs need to be converted into an omni-directional PDPs, which encapsulate all the information from directional PDPs (144 in our case) in an organized manner. Some work has been done on the topic of synthesizing omni-PDPs from directional PDPs in \cite{Hur2014SynchronousCS,omnipdpcitation}. We use the method discussed in \cite{Hur2014SynchronousCS} to serve the purpose. If $X(\tau,n,j)$ is the directional PDP with time index $\tau$ and in PAC $n$ for the $j$-th scan, the omni-PDP $S(\tau,j)$ can be simply calculated using the equation
\beq
S(\tau,j)=\max\limits_{n} X(\tau,n,j),
\eeq
Note that the dimensions of $S(\tau,j)$ are $N_{dly} \times N_{scans}$.

\section{LOS Analysis}
\label{section:LOSAnalysis}
\subsection{AoA/AoD Estimation}
In this subsection, we will propose a method of extracting the AoA/AoD pair of the LOS path from the measurement data from all 12 cases. We use Received Signal Strength Indicator (RSSI) in all pointing angle combinations (PAC) as the parameter to be compared. We firstly identify the sample that contains the LOS component from the synthesized omni-PDP. The LOS component will be the first to arrive at RX (since it is the shortest path) so the time at which the first peak is observed at the receiver is the LOS component. There will be $N_{scan}$ number of RSSIs present in measurement data for each sample. The overall $RSSI_{los}$ is extracted by averaging RSSI from the PDP sample containing LOS component across all these scans. Mathematically, if $X$ is the measurement data tensor with dimensions described in section \ref{sec:MeasurementSetup}, $k_{los}$ is the LOS component index, the $RSSI_{LOS}$ is given by:
\beq
\label{eq:y}
 RSSI_{LOS}(n)= \frac{1}{N_{scans}} \sum_{j=1}^{N_{scans}} X(k_{los},n,j),
\eeq
Now that the $RSSI_{LOS}$ from measurement is obtained, we perform a \textit{least squares (LS) direction finding} procedure to find out the AoA/AoD  pair for the LOS component. 
Suppose $\omega = (\phi_{TX},\phi_{RX},\theta_{TX},\theta_{RX})$ is a possible AoA / AoD.  
Then, we would expect,
\beq
     RSSI_{LOS}(n) \approx RSSI_0 + G(n,\omega),
\eeq
where $RSSI_0$ is the RX power of the path before the beamforming gain and $G(n,\omega)$ is the antenna gain for the true AoA/AoD angle $\omega$.  
To determine $RSSI_0$ and $\omega$, we solve,
\beq
    \argmin_{RSSI_0, \omega} \sum_n |RSSI_{LOS}(n) - RSSI_0 - G(n,\omega)|^2,
\eeq
Performing the minimization over $RSSI_0$, we obtain:
\beq
\label{eq:angle_est}
  \widehat{\omega} = \argmin_{\omega}  \mathrm{var}( RSSI_{LOS}(n)  - G(n,\omega)), 
\eeq
where $\mathrm{var}(X(n))$ is the 
variance over $n$.
This procedure provides an 
estimate for the AoA/AoD pair
in each case.  The angles relative
to a fixed reference can then be
computed from Table \ref{table:antorientation} 
using simple geometry.

\subsection{Comparison to ray-tracer}
To verify the AoA/AoD estimates for all cases, we use a commercial ray-tracing simulator called Scenargie by STE \cite{scenargie}. 
The TX and RX antenna patterns are loaded into the ray-tracer. 
Then ray-tracing is executed for the LOS path only (we discuss NLOS later). After ray-tracing is complete, we get an estimated RSSI at the output of the simulator $RSSI_{est}(n)$ for each
PAC $n$ in each case.   
Fig. \ref{fig:losvalidation} shows
the comparison between the estimated
RSSI from the ray tracer
$RSSI_{est}(n)$ and our measurements $RSSI_{los}(n)$.
We see that the two estimates 
are often within a few dB, signifying that there is a close correspondence between the measurements and the ray-tracer.
For each of the 12 cases,
we have also computed
the correlation coefficient,
\beq 
 \rho = \frac{cov(RSSI_{los},RSSI_{est})}{\sigma_{RSSI_{los}}\sigma_{RSSI_{est}}},
 \eeq
where the covariance and standard
deviations are computed over the
PACs in that use case.
Values of $\rho \geq 0.85$
are found in 10 out of 12 cases. This means that the proposed procedure for estimating AoA/AoD pair for LOS was successful and agrees with commercial
ray tracer.

\subsection{LOS power analysis}
\begin{figure}[t!]
 \centering
   
       \includegraphics[scale=0.4]{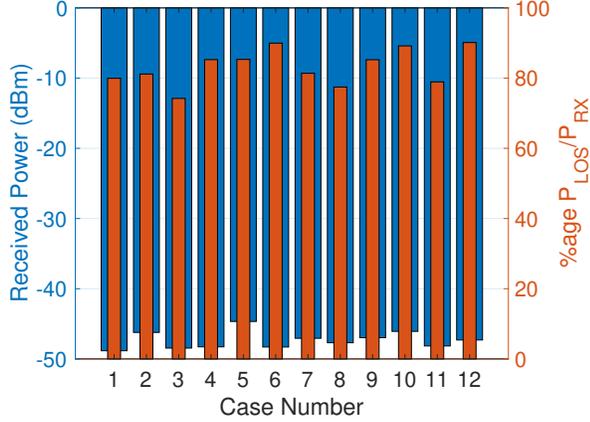}

\caption{Total recived power on LOS path $P_{LOS}$(blue) and percentage of power LOS contributes to the total received power $P_{RX}$ (red)}
\label{fig:lospowerfig}
\end{figure}

In the next step of LOS analysis, we calculate the power in LOS components $P_{LOS}$ in all the cases, which is done by averaging the power in LOS component from the synthesized omni-PDP $S(t,j)$ over all scans, and compare it to the total received power. The total received power $P_{RX}$ is calculated from the synthesized omni-PDP  in two steps. We first calculate average power $P_{av}$ in each sample from omni-PDP by averaging over number of scans. Mathematically,
\beq
P_{av}(\tau) = \frac{1}{N_{scans}} \sum_{j=1}^{N_{scans}} S(\tau,j),  
\eeq
The power contained in LOS path is simply the power at the LOS index of $P_{av}(\tau) $. The noise power $P_N$ is then subtracted from the total power to get the total received power. $P_N$ is calculated based on the assumption that anything received earlier than LOS component in PDP is noise. This is a fair assumption since LOS component is the shortest path between the TX and RX, so any signal received before that is essentially noise. If $k_{los}$ is the LOS sample in the averaged omni-PDP $P_{av}(\tau) $, the noise power can be calculated using 
\beq
P_N = \frac{1}{k_{los}-M} \sum_{i=1}^{k_{los}-M} P_{av}(i),
\eeq
where $M$ is chosen to ensure that the width of the LOS components' pulse is not taken into account, while calculating noise power. After calculating $P_N$, the received signal power $P_{RX} =  P_{av}(\tau) -P_N$ and then the LOS link power $P_{LOS}$=$P_{av}(k_{los})$ are calculated. With $P_{LOS}$, $P_N$ and $P_{RX}$, we can see how the received power on LOS link in each case looks like. Apart from that, we calculate percentage of total received power is obtained from LOS. The results have been summarized in Fig. \ref{fig:lospowerfig} . It can be observed from Fig. \ref{fig:lospowerfig} that LOS path is responsible for most percentage of the total power varying from 79 to 90 percent with an average of 83 percent in all cases. The RSSI on the LOS link fluctuates between -44 and -49 dBm with an average of around -47.33 dBm across all cases. 

\section{NLOS Analysis}
\label{sec:Numerical Results}
\subsection{AoA/AoD Estimation}
As seen in section \ref{section:LOSAnalysis}, LOS path accounts for a major percentage of the total power for mmWave system, but the LOS component will not always be available because of the higher susceptibility of the signals to be blocked \cite{rappaport2014millimeter}. We analyze the NLOS paths in the measurement venue to see what options the system have as an alternative to LOS path.
\begin{figure}[b!]
 \centering
   
       \includegraphics[scale=0.4]{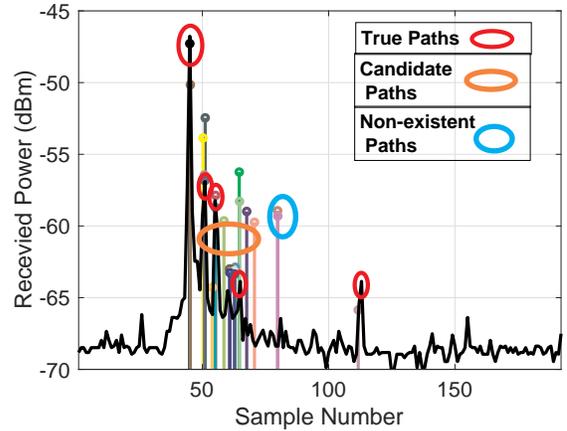}

\caption{Omni PDP from measurement case 8 (black) vs PDP generated by ray-tracer (stem)}
\label{fig:PDPcomp}
\end{figure}
The first part is to identify NLOS paths from the measurement data. This is done by studying the omni-PDPs of all the cases. All the peaks above a certain threshold are noted. The threshold for every case is chosen to be $P_N$ + $3\textrm{dB}$. A total of 6 peaks including the LOS path was observed across all cases.

After recording all the NLOS peaks, we extract the RSSI from the measurement data for all the NLOS path in question, called $RSSI_{nlos}$ from the current case. It should be noted that the choice of case is also critical to examine the path route for example in the cases with rotations in azimuths, the paths arriving from the sides will be dominant as compared to up-tilts and down-tilts, where, the paths from the ceiling and floor will be dominant.  The method for extracting is similar to how we extract RSSI in LOS case. The difference is sample number ($k_{los}$ will now be changed to the delay index associated with each particular NLOS path).

Estimating the AoA/AoD for each NLOS path using the \textit{LS direction finding} method from (\ref{eq:angle_est}) is next. We estimate AoA/AoD  ($\widehat{\omega}_{nlos}$) for each NLOS path in every case. This estimate will then help us to trace the physical location of the path. We proceed to the next step after the estimation is complete,
  
\subsection{Finding Physical Locations of NLOS Paths}

The \textit{LS direction finding} provides the directions of various NLOS multipath components.
To validate the estimates, we again
use ray tracing to confirm
that the estimated paths correspond
to actual expected paths in the environment.
We use the identical ray-tracer as in Section \ref{section:LOSAnalysis}). The ray-tracer is configured to show the paths with a maximum of 2 reflections, 1 transmission and 0 diffractions. As described in \cite{article}\cite{Jarvelainen2016IndoorPC}\cite{scattering}, mmWave systems will mostly rely on reflections for multi-path propagation, justifying the choice to focus mostly on reflections. We expect some transmissions since there are glass panels on cubicle walls. After ray-tracing is performed, all the peaks observed in the PDP from ray-tracer and measurements are compared.

We define three types of paths here, which will be used in the upcoming sections. \textit{Candidate Paths} are the paths that can be observed in the PDP obtained from the ray-tracer, and the delay timing of these paths is close with those found in measurements. \textit{True paths} are paths suggested by the ray-tracer that can be confirmed by the measurement data. \textit{Non-existent paths} are the ones that appear in the PDPs generated by ray-tracer but are not present in the measurement PDPs. Note that there will be more than one candidate path from the ray-tracer, because it is possible that paths with different routes but similar path lengths exist. The temporal information from the omni-PDP obtained from measurements and the spatial information estimated from \textit{LS direction finding} are key to shortlisting candidate paths. Fig. \ref{fig:PDPcomp} shows PDP generated by ray-tracer and synthesized. Omni-PDP for case 8, where all of the aforementioned paths can be observed.
\begin{figure}[t]
 \centering
   
       \includegraphics[scale=0.4]{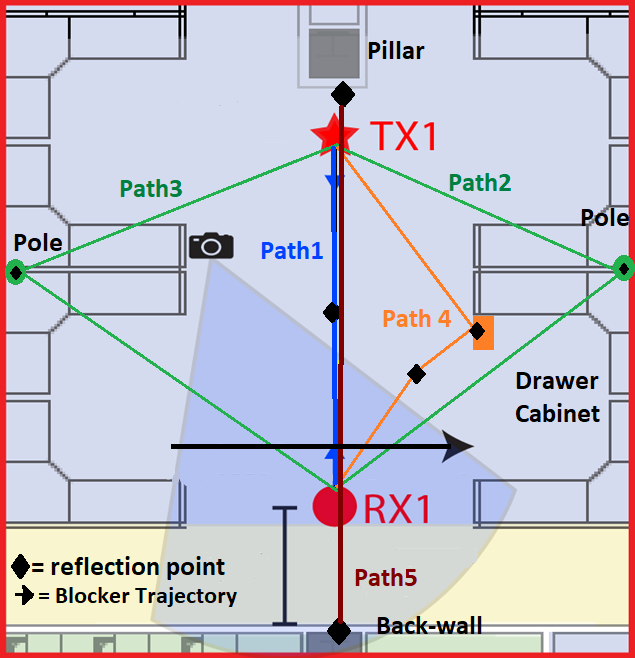}

\caption{Top view of the venue with determined \textit{true paths} and blocker trajectory for section \ref{sec:blockage}}

\label{fig:refpaths}
\end{figure}

\begin{figure}[t!]
 \centering
   
       \includegraphics[trim= {0cm 0cm 0cm 0cm},clip,width=0.4\textwidth]{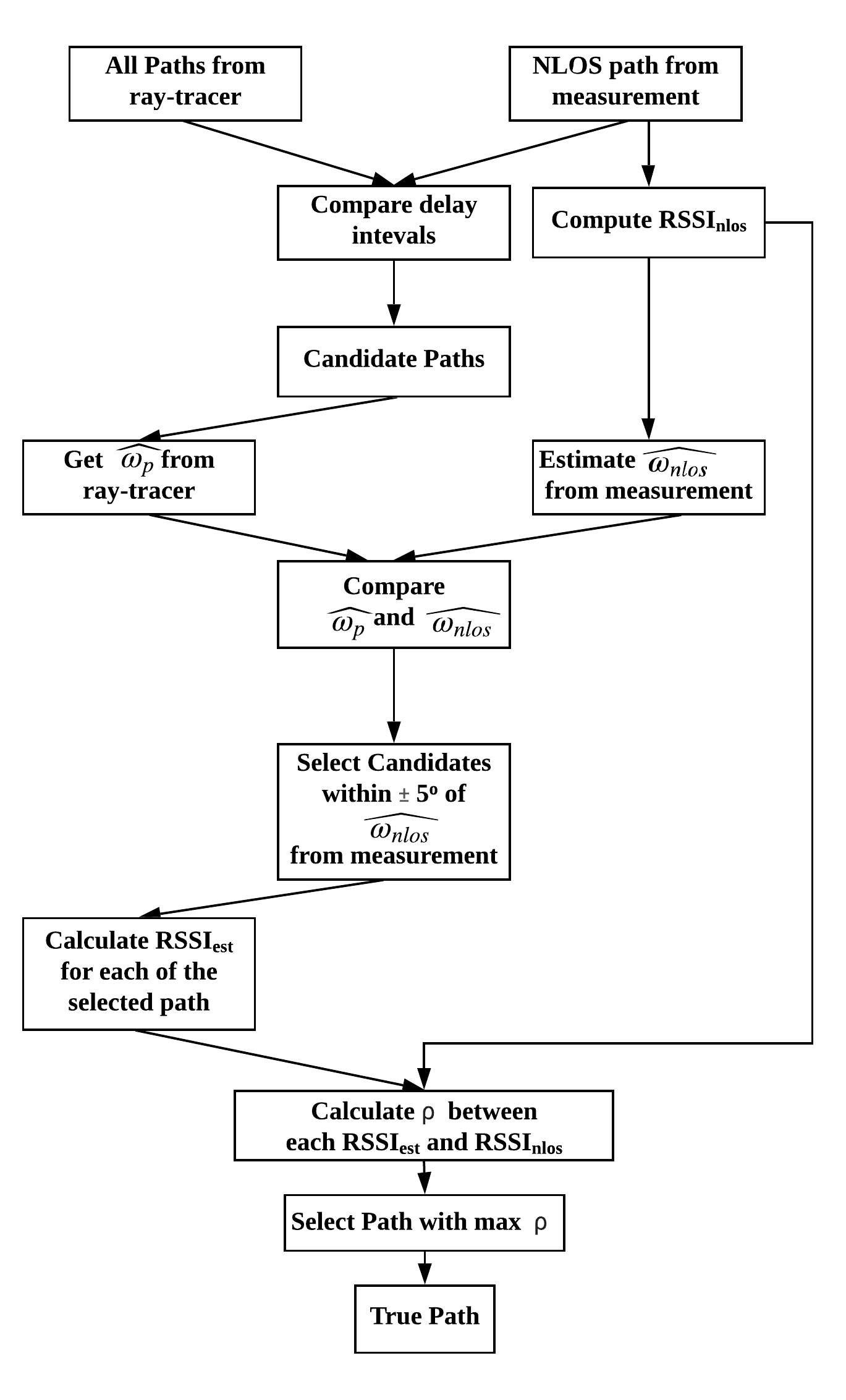}

\caption{Flow chart for selection of true path from candidate paths}
\label{fig:flowchart}
\end{figure}
\begin{figure*}[t!]
 \centering
   
       \includegraphics[trim= {2cm 3.5cm 0cm 0cm},clip,width=0.9\textwidth]{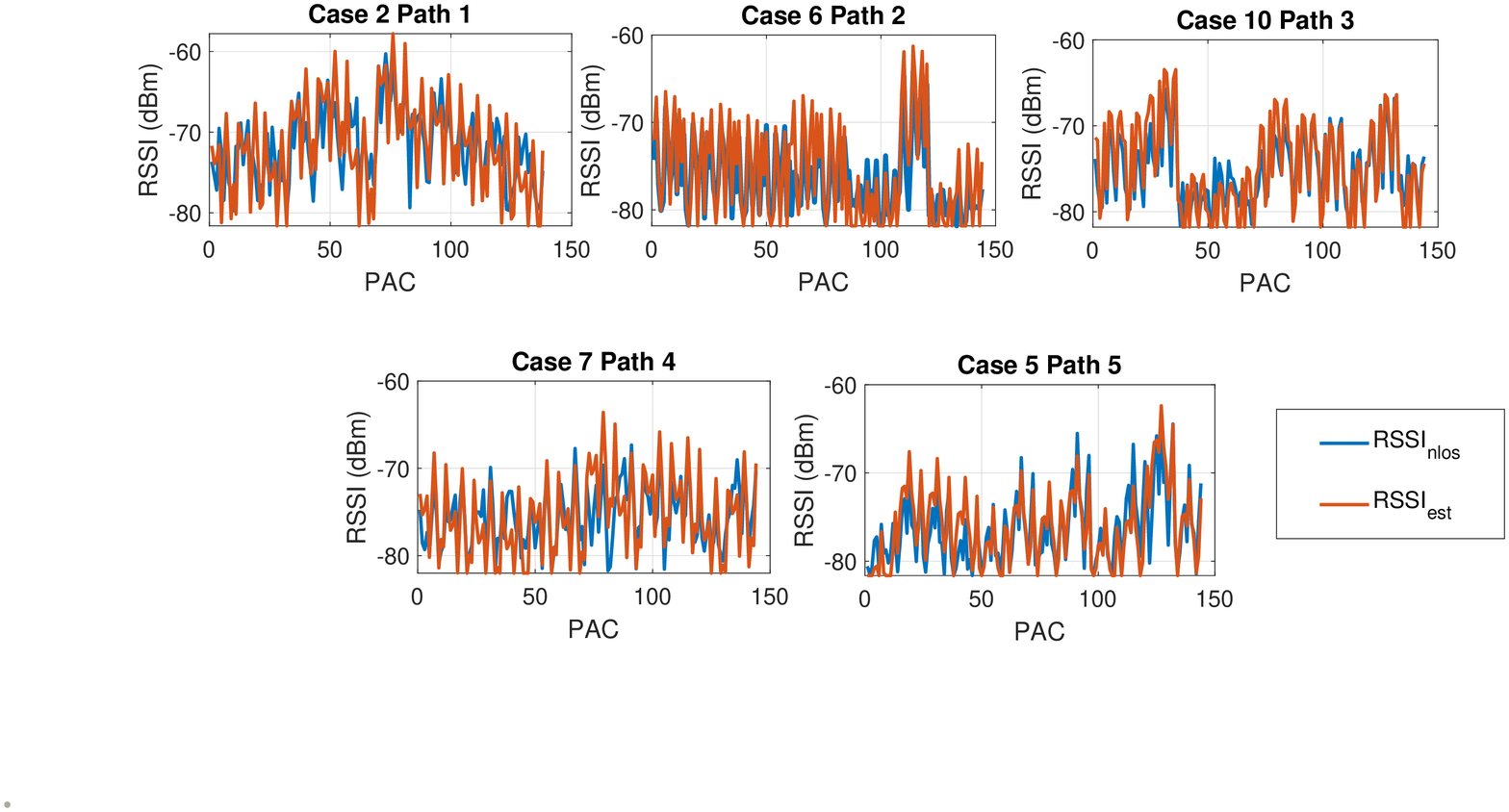}

\caption{$RSSI_{nlos}$ vs $RSSI_{est}$ of identified paths for the cases with best $\rho$}
\label{fig:nlosvalidation}
\end{figure*}

 In the next step, we extract \textit{true paths} from the short-listed candidate paths. The procedure for selecting the \textit{true paths} has the following steps. The \textit{candidate paths} have similar temporal information as the NLOS paths in measurement data (discarding non-existent paths). We use the AoA/AoD estimates  $\widehat{\omega_{nlos}}$ to further narrow down the search for the \textit{true path}. We note the AoA/AoD pairs for the $pth$ candidate path i.e. $ \widehat{\omega_{p}}= (\theta_{TXp},\theta_{RXp},\phi_{TXp},\phi_{RXp})$.  The AoA/AoD pairs from ray-tracer that are closest to the $\widehat{\omega_{nlos}}$ (within $\pm 5^o$ to accommodate for alignment and dimension measurement error)  are chosen for further analysis. $RSSI_{est}$ is calculated on all these paths and compared with respective $RSSI_{nlos}$. The path with the highest $\rho$ is declared to be the true path. The procedure for extraction of true paths from candidate paths is shown in  Fig.\ref{fig:flowchart}.

The identified paths other than LOS are listed below:
\begin{itemize}

    \item $Path 1$: The floor (First order reflection)
    \item $Path 2$ $\&$  $Path 3$: The poles supporting the cubicle jungle (First order reflection, 2 symmetrical paths)
    \item $Path 4$: Drawer cabinet and the floor (Second order reflection)
    \item $Path 5$: Back-wall and the pillar (Second order reflection) 
\end{itemize}

Six out of seven dominant paths (including LOS) were identified and verified from the measurements using the proposed procedure. The path difference between LOS and the reflection from ceiling is less than 24 cm (observed from the ray-tracer) so these paths cannot be resolved by the sampling rate ($1.25 GS/s$) of the system. LOS however will dominate since it has no interaction with the environment. Two paths are symmetric meaning they arrive at the RX at the same time and are visible when the TX/RX is rotated towards them.

Fig. \ref{fig:refpaths} shows a diagram of the measurement environment and the path locations. Figure \ref{fig:nlosvalidation} shows how $RSSI_{nlos}$ and $RSSI_{est}$ compare after going through the selection procedure. Only the cases with highest $\rho$ are shown for each path identified. A clear correspondence between $RSSI_{nlos}$ and $RSSI_{est}$ show that we were successful in extracting the \textit{true paths} from the candidates using the proposed method. 
\begin{table}[]
\begin{tabular}{|l|c|c|c|c|}
\hline
\textbf{Cases} & \multicolumn{1}{l|}{\textbf{\begin{tabular}[c]{@{}l@{}}Path 1\\ $P_{nlos}$\\ (P')\\ dBm\\ PL=1.8 dB\\ RL=5 dB\end{tabular}}} & \multicolumn{1}{l|}{\textbf{\begin{tabular}[c]{@{}l@{}}Path 2 or 3\\ $P_{nlos}$\\ (P')\\ dBm\\ PL=6.1 dB\\ RL=9.8 dB\end{tabular}}} & \multicolumn{1}{l|}{\textbf{\begin{tabular}[c]{@{}l@{}}Path 4\\ $P_{nlos}$\\ (P')\\ dBm\\ PL=3.5 dB\\ RL=8.5 dB\end{tabular}}} & \multicolumn{1}{l|}{\textbf{\begin{tabular}[c]{@{}l@{}}Path 5\\ $P_{nlos}$\\ (P')\\ dBm\\ PL=12.7 dB\\ RL= 6.5 dB\end{tabular}}} \\ \hline
\textbf{1}     & \begin{tabular}[c]{@{}c@{}}-58.51\\ (-9.68)\end{tabular}                                                                            & -                                                                                                                                          & \begin{tabular}[c]{@{}c@{}}-66.68\\ (-17.84)\end{tabular}                                                                             & \begin{tabular}[c]{@{}c@{}}-67.77\\ (-18.94)\end{tabular}                                                                               \\ \hline
\textbf{2}     & \begin{tabular}[c]{@{}c@{}}-60.22\\ (-14.00)\end{tabular}                                                                           & \begin{tabular}[c]{@{}c@{}}-66.68\\ (-20.45)\end{tabular}                                                                                  & \begin{tabular}[c]{@{}c@{}}-58.97\\ (-12.75)\end{tabular}                                                                             & \begin{tabular}[c]{@{}c@{}}-67.02\\ (-20.79)\end{tabular}                                                                               \\ \hline
\textbf{3}     & \begin{tabular}[c]{@{}c@{}}-56.70\\ (-8.24)\end{tabular}                                                                            & \begin{tabular}[c]{@{}c@{}}-59.20\\ (-10.75)\end{tabular}                                                                                  & -                                                                                                                                     & \begin{tabular}[c]{@{}c@{}}-67.74\\ (-19.29)\end{tabular}                                                                               \\ \hline
\textbf{4}     & \begin{tabular}[c]{@{}c@{}}-58.67\\ (-10.39)\end{tabular}                                                                           & -                                                                                                                                          & -                                                                                                                                     & \begin{tabular}[c]{@{}c@{}}-66.82\\ (-18.54)\end{tabular}                                                                               \\ \hline
\textbf{5}     & \begin{tabular}[c]{@{}c@{}}-60.70\\ (-16.04)\end{tabular}                                                                           & -                                                                                                                                          & \begin{tabular}[c]{@{}c@{}}-62.37\\ (-17.71)\end{tabular}                                                                             & \begin{tabular}[c]{@{}c@{}}-63.03\\ (-18.37)\end{tabular}                                                                               \\ \hline
\textbf{6}     & \begin{tabular}[c]{@{}c@{}}-60.51\\ (-12.21)\end{tabular}                                                                           & \begin{tabular}[c]{@{}c@{}}-63.52\\ (-15.22)\end{tabular}                                                                                  & -                                                                                                                                     & \begin{tabular}[c]{@{}c@{}}-69.67\\ (-21.37)\end{tabular}                                                                               \\ \hline
\textbf{7}     & \begin{tabular}[c]{@{}c@{}}-57.48\\ (-10.43)\end{tabular}                                                                           & \begin{tabular}[c]{@{}c@{}}-64.52\\ (-17.47)\end{tabular}                                                                                  & \begin{tabular}[c]{@{}c@{}}-60.86\\ (-13.81)\end{tabular}                                                                             & \begin{tabular}[c]{@{}c@{}}-67.90\\ (-20.85)\end{tabular}                                                                               \\ \hline
\textbf{8}     & \begin{tabular}[c]{@{}c@{}}-58.63\\ (-10.94)\end{tabular}                                                                           & \begin{tabular}[c]{@{}c@{}}-67.42\\ (-19.73)\end{tabular}                                                                                  & \begin{tabular}[c]{@{}c@{}}-60.21\\ (-12.52)\end{tabular}                                                                             & \begin{tabular}[c]{@{}c@{}}-67.51\\ (-19.82)\end{tabular}                                                                               \\ \hline
\textbf{9}     & \begin{tabular}[c]{@{}c@{}}-56.71\\ (-9.74)\end{tabular}                                                                            & \begin{tabular}[c]{@{}c@{}}-66.11\\ (-19.14)\end{tabular}                                                                                  & -                                                                                                                                     & \begin{tabular}[c]{@{}c@{}}-68.12\\ (-21.15)\end{tabular}                                                                               \\ \hline
\textbf{10}    & \begin{tabular}[c]{@{}c@{}}-61.55\\ (-15.48)\end{tabular}                                                                           & \begin{tabular}[c]{@{}c@{}}-64.60\\ (-18.53)\end{tabular}                                                                                  & \begin{tabular}[c]{@{}c@{}}-61.64\\ (-15.57)\end{tabular}                                                                             & \begin{tabular}[c]{@{}c@{}}-66.09\\ (-20.02)\end{tabular}                                                                               \\ \hline
\textbf{11}    & \begin{tabular}[c]{@{}c@{}}-62.54\\ (-14.39)\end{tabular}                                                                           & \begin{tabular}[c]{@{}c@{}}-66.31\\ (-18.16)\end{tabular}                                                                                  & \begin{tabular}[c]{@{}c@{}}-63.94\\ (-15.79)\end{tabular}                                                                             & \begin{tabular}[c]{@{}c@{}}-66.01\\ (-17.86)\end{tabular}                                                                               \\ \hline
\textbf{12}    & \begin{tabular}[c]{@{}c@{}}-62.11\\ (-14.81)\end{tabular}                                                                           & \begin{tabular}[c]{@{}c@{}}-66.90\\ (-19.60)\end{tabular}                                                                                  & -                                                                                                                                     & \begin{tabular}[c]{@{}c@{}}-68.25\\ (-20.95)\end{tabular}                                                                               \\ \hline
\end{tabular}

\caption{NLOS Power Analysis}
\label{table:nlospower}
\end{table}

\subsection{NLOS Power Analysis}
After the identification of the NLOS paths. We continue to do the analysis by investigating power of the NLOS $P_{NLOS}$ paths similar to section \ref{section:LOSAnalysis}. The procedure is the same as the LOS path, the only thing different is the sample to be extracted from the synthesized omni-PDP. The noise calculation is the same as the LOS analysis. The data for NLOS paths for all cases can be seen in table \ref{table:nlospower}. $PL$ refers to path loss suffered by the NLOS signal $PL_{nlos}$ compared to the pathloss suffered by the LOS signal $PL_{los}$ i.e $PL$=$PL_{nlos}-PL_{los},$ and $RL$ means reflection loss because of the reflection(s). If the path in question is absent in the case, it is marked by '-'. The entry on top in each row is the link $P_{nlos}$ and the bottom entry (in parenthesis) is power with respect to the LOS component in that case ($P'$=$\frac{P_{nlos}}{P_{los}}$).Table \ref{table:nlospower} shows that $Path1$ is the strongest in most of the cases owing to the minimum $PL$, but it can be observed that for some instances, for example case 2, $Path4$ provides a better link than $Path1$ even though the signal suffers a double reflection. Although it only happens once (case 2), this is an interesting finding, which tells us that we can even rely on the second order reflections in mmWave systems. This phenomenon is an artifact of the beamforming deployed at TX/RX. $Path5$ is the weakest one in all cases because of the $PL$ and two reflections. It can also be observed from table \ref{table:nlospower} that the power in NLOS component is 9 to 20 dB weaker than the power in LOS component.
\section{Blockage Analysis}
\label{sec:blockage}
\begin{figure}[t!]
 \centering
   
       \includegraphics[trim={1cm 2cm 0cm 0cm},clip,width=0.4\textwidth]{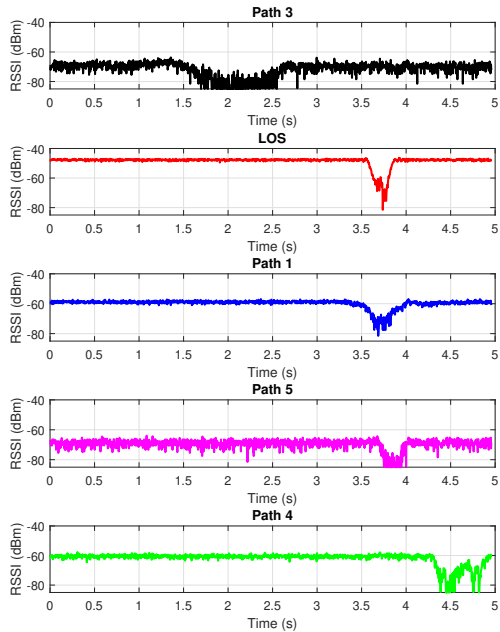}

\caption{Evolution of RSSI on the best PAC, with time on each path with blockage event}
\label{fig:blockage}
\end{figure}

With a complete understanding of the static nature of this channel, we now move to a brief discussion of dynamic blockage. We analyze the behavior by picking case 8, since it has the most number of identified paths present. The measurement setup for case is the same as mentioned in table \ref{table:antorientation}. The difference is that a human blocker is moving between the TX and RX with the trajectory shown in Fig.\ref{fig:refpaths}. We analyze the evolution of the RSSI for strongest PAC, with time. Different paths will have different best PACs depending on angles of arrival and departure. The time evolution of the RSSI on each path with blockage events can be seen from the Fig.\ref{fig:blockage}. $Path3$ gets blocked first since it is the first to encounter the blockage event, LOS $Path1$, $Path5$ are blocked simultaneously since there is no horizontal displacement for either of these paths. And  $Path4$ is blocked at the end, since it is the last path to suffer blockage. $Path2$ is not clearly visible in the PDP. The analysis of the sequence of blockages further confirms the correctness of the identified paths. Another takeaway from the blockage analysis is that at least two paths are available for communication at any point in time.  This analysis helps us build robust systems operating at mmWaves even with blockage.
\section{Conclusion}
\label{sec:concl}
Measurements of the LOS and NLOS paths propagation characteristics are critical for the design of systems operating at mmWave frequencies. 
In this work, we provide a direction estimation procedure for LOS and NLOS paths from a phased-array measurement system.
The results are tested in an office
scenario and agree closely with simulations from a ray-tracer programmed
with a model for the room. 
In this particular office, it is shown that that the observed LOS component accounts for about 83 percent of the total power with a maximum of 90 percent and a minimum of 79. The observable NLOS paths are at least 9 dB weaker than the LOS path with the worst case being 20 dB. NLOS components that undergo second order reflections can also be extracted and used for communication purposes. We conclude with the analysis of the identified paths under the influence of a human blockage that at least two paths are available for communication at any point in time. An analysis of the blockage event timing and blocker motion agrees with the expected blockage timeline for the identified NLOS paths.

\section*{Acknowledgements}
This work was supported in part by 
the Sony corporation, NSF grants
1936332, 1824434, 1833666, 1564142; NYU WIRELESS and its industrial affiliates;
NIST grant 70NANB17H166; and the SRC.

\balance
\bibliographystyle{IEEEtran}
\bibliography{references}
\end{document}